\documentclass[10pt,aip,apl,twocolumn,showpacs,floatfix,superscriptaddress]{revtex4-1}
\usepackage{amsmath,graphics,epsfig,mathrsfs,bm}

\newcommand{\ti}{\textit}

\newcommand{\alp}{\alpha}

\newcommand{\etal}{{\em et al.~}}

\newcommand{\Cp}{C$_{5}$H$_{5}$}
\newcommand{\Bz}{C$_{6}$H$_{6}$}
\newcommand{\Cot}{C$_{8}$H$_{8}$}

\begin{document}
\title{Peculiarities of Spin Polarization Inversion at a Thiophene/Cobalt Interface}
\author{Xuhui Wang, Zhiyong Zhu, Aurelien Manchon, Udo Schwingenschl\"ogl}
\email{udo.schwingenschlogl@kaust.edu.sa,00966(0)544700080}
\affiliation{King Abdullah University of Science and Technology
(KAUST), Physical Science and Engineering Division, Thuwal
23955-6900, Saudi Arabia}
\date{\today}

\begin{abstract}
We perform \ti{ab-initio} calculations to investigate the spin
polarization at the interface between a thiophene molecule and
cobalt substrate. We find that the reduced symmetry in the
presence of a sulfur atom (in the thiophene molecule) leads to a
strong spatial dependence of the spin polarization of the
molecule: The two carbon atoms far from the sulfur acquire a
polarization opposite to that of the substrate, while the carbon
atoms bonded directly to sulfur possess the same polarization as
the substrate. We determine the origin of this peculiar spin
interface property as well as its impact on the spin transport.
\end{abstract}

\maketitle

The discovery of spin-related phenomena in organic materials
promises exciting development towards low cost yet efficient
spintronic devices \cite{dediu-ssc-2002,xiong-nature-2004}. In the
focus is the study of properties combining traditional
inorganic and the organic materials, such as the interface between
an organic molecule and a ferromagnetic metal, the so-called
\ti{spinterface} \cite{sanvito-2010,new1,new2,new3}. A key
ingredient to spin manipulation in such a hybrid structure is spin
injection, which continuously demands higher efficiency and better
understanding
\cite{drew-natmater-2009,cin-natmater-2009,barraud-natphys-2009}.
Therefore, the behavior of the spin at a spinterface is of great
importance to unravel the mysteries of spin transport in hybrid
systems. Experimentally, spin-polarized scanning tunneling
microscopy \cite{spstm-science-2007, iacovita-prl-2008,
brede-prl-2010, methfessel-2011} and X-ray absorption spectroscopy
\cite{wende-natmat-2007} have been shown to be useful tools to
provide detailed spin information at a surface. Despite numerous
experiments, not many theoretical efforts have been put forward to
scrutinize the uniqueness of a spinterface, where the level of
complexity places a challenge to simple analytical models. On the
other hand, \ti{ab-initio} calculations provide a well established
platform to deepen the understanding of complex systems.

Recently, for a group of carefully selected molecules of similar
electronic properties, Atodiresei \etal have carried out an
\ti{ab-initio} study towards the interface formed by organic
molecules adsorbed on a ferromagnetic Fe/W(110) surface
\cite{atodiresei-prl-2010}. A polarization inversion on the
molecular side has been found, which is consistent with
experimental results using spin-polarized scanning tunneling
microscopy: the sign of the spin polarization of the molecule is
opposite to that of the substrate to which it is chemically
bonded. The selection of the molecules is based on the
$\pi(p_{z})$ electron systems which usually have a strong
interaction between the $\pi$ electrons of the molecule and the
$d$ orbitals of the ferromagnetic substrate, thus allowing a
Zener-type exchange interaction (between the $p_{z}$ and $d$
orbitals) to trigger the polarization inversion. The selected
molecules (\Cp, \Bz, and \Cot), from the molecular structure point
of view, are highly symmetric in the molecular plane and composed
of only carbon (C) and hydrogen (H) atoms. For example, the \Cp\
molecule is a unit of an even larger molecule as compared to
H$_{2}$Pc.

In this Letter, we report a peculiar picture of spin
polarization inversion as a result of asymmetry in the molecular structure on the
functionality of $p_{z}$-$d$ exchange.
We perform \ti{ab-initio} calculations to investigate
an organic-ferromagnetic metal interface formed by a thiophene (C$_{4}$H$_{4}$S) molecule
adsorbed on a cobalt (Co) substrate. The motivations for this choice are as follows.
Besides serving as the basis unit to form polythiophene
(such as $\alp$-sexithiophene, commonly known as T6 molecule)
that is widely used in organic spintronics \cite{dediu-ssc-2002},
the thiophene molecule is less symmetric than \Cp\ since one
C is substituted by an S atom. The impact of the reduced
symmetry on the local spin polarization at a spinterface
is yet to be explored. A small molecule with a relatively simple composition,
like thiophene, gives us the advantage to concentrate on the
significance of a particular critical mechanism,
such as the Zener exchange scrutinized in this paper.
Moreover, the delocalized electron pairs of sulfur
in the $\pi$ electron system render the reactivity of thiophene to be comparable (though less)
to molecules like \Cp. Moreover, a T6 molecule consists of six thiophenes bonded
linearly through the 2- and 5-positions.
We expect the spin interface properties associated with a thiophene molecule
to persist into the chain, thus having direct implications for T6 molecules.

\begin{figure}[t]
\centering
\includegraphics[clip,width=0.5\textwidth]{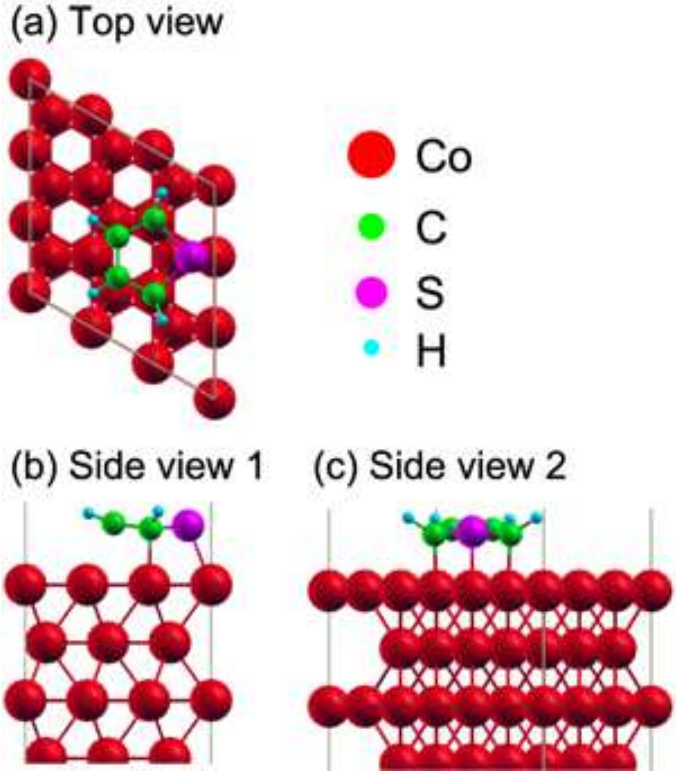}
\caption{(color online). A thiophene molecule adsorbed on a
Co(001) surface after structure relaxation.} \label{fig:struct}
\end{figure}

The structure is shown in Fig.\ref{fig:struct}. A thiophene molecule is placed on the Co(001)
surface. Before the structure relaxation, the initial position of the
molecule on the Co surface is selected to be as symmetric as possible with respect to the S
atom to minimize the total force. We next perform a charge relaxation to allow the force
exerted on the molecule to be minimal.
As shown in Figs.\ \ref{fig:struct}(c) and (d), after the structure relaxation
the molecule plane is no longer flat: the S-Co bond
becomes longer than in its initial configuration.
Full-potential linearized augmented plane wave calculations are
performed using the WIEN2k package \cite{wien2k}. To simulate the Co(001)
surface, we employ a slab geometry with a $3\times 3$ in-plane surface
unit cell and eight Co atomic layers that are terminated by a vacuum layer of
18 \AA. A threshold energy of $-6.0$ Ry, which separates valence and
core states, is used in all calculations. We employ for the muffin-tin radius
$R_{\text{mt}}$ values of 2.2 bohr, 1.2 bohr, 1.5 bohr and 0.7 bohr for Co, C, S and H,
respectively. A small value of $R_{\text{mt}}K_{\text{max}}=2$
is used due to the H atoms in the system \cite{wien2k}. The
$\vec{k}$-mesh is set to be $6\times 6\times 1$ and all calculations are
conducted with the Perdew-Burke-Ernzerhof parametrization of the
generalized gradient approximation to the exchange correlation potential \cite{pbe}.
Force convergence of 5 mRy/bohr is employed in the structure relaxation.
In systems consisting of macromolecules (such as CoPc) \cite{brede-prl-2010},
long range van der Waals forces can be relevant for the relaxation and, thus, for
the spin polarization. However, in the present setup they are not decisive and will
be omitted in the calculations. We find a strong chemical bonding
between the molecule and the substrate, which dominates the van der Waals forces
by several orders of magnitude. Similar findings have been reported for related
molecules \cite{atodiresei-prl-2010}.

\begin{figure}[t]
\centering
\includegraphics[clip,width=0.5\textwidth]{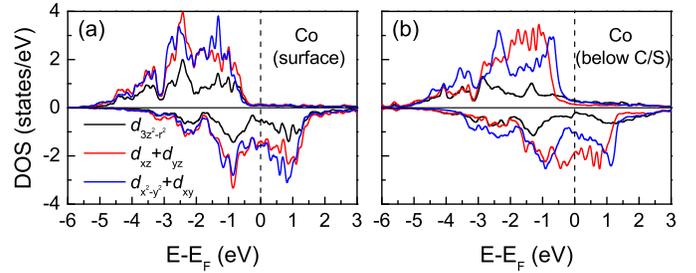}
\caption{(color online). Density of states (a) of a Co atom on a
clean surface and (b) averaged over the 4 Co atoms below the C/S
atoms of the thiophene molecule.} \label{fig:dos-co}
\end{figure}

For the purpose of comparison, we show in Fig.\
\ref{fig:dos-co}(a) the spin-polarized local density of states of
a Co atom in a clean surface. Figure \ref{fig:dos-co}(b) depicts
the density of states averaged over the 4 Co atoms below the
thiophene molecule. By comparing panels (a) and (b), the
out-of-plane $d_{3z^{2}-r^2}$, $d_{xz}$, and $d_{yz}$ orbitals of
Co are modified due to the coupling to the $p_{z}$ orbital of C.
The in-plane $d_{x^{2}-y^{2}}$ and $d_{xy}$ orbitals are much less
affected. This suggests that the hybridization scheme in the
present case is similar to the interface between Fe and the big
$\pi(p_{z})$ molecules \cite{atodiresei-prl-2010}. Being a
neighbor to Fe ($3d^{6}4s^{2}$) in the periodic table, the
electron configuration of Co is $3d^{7}4s^{2}$. Figure
\ref{fig:dos-co} also shows that the change is much less dramatic
for the spin-up states than for the spin-down states. Therefore,
we expect that the hybridization scheme exhibited in Fig.\
\ref{fig:dos-co} accommodates a $p_{z}$-$d$ Zener exchange
interaction that eventually gives rise to a spin polarization
inversion on the molecule.

\begin{figure}[t]
\centering
\includegraphics[clip,width=0.5\textwidth]{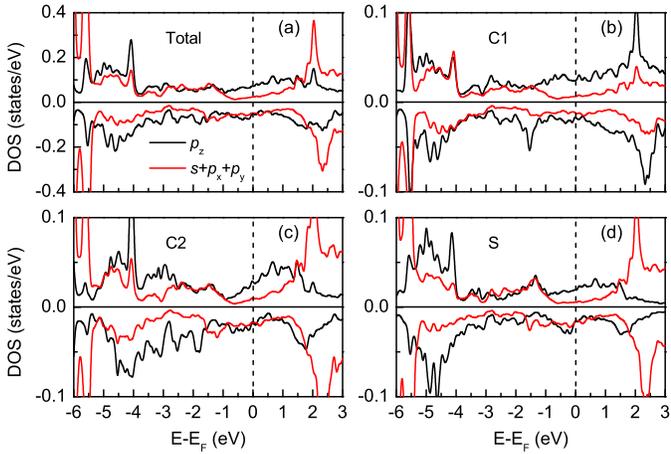}
\caption{(color online). Density of states of (a) the entire
thiophene on the Co(001) surface, (b) the two C atoms at the 3-
and 4-positions (C1 group), (c) the two C atoms at the 2- and
5-positions (adjacent to the S atom; C2 group), and (d) the S
atom.} \label{fig:dos-thiophene}
\end{figure}

For the molecule, Fig.\ \ref{fig:dos-thiophene} shows the density of states
of the entire molecule in panel (a), of the C atoms in panels (b) and (c), and
of the S atom in panel (d). In contrast to a \Cp\ molecule,
the reduced symmetry of thiophene motivates us to divide the C atoms
into two groups: in group C2 (the so-called 2- and 5-positions)
the C atoms are adjacent to S, while group C1 comprises the other atoms (3- and 4-positions).
At the Fermi energy, the out-of-plane $p_{z}$ orbital exhibits a very small spin imbalance as
the spin up state slightly dominate. In contrast, on the Co site the spin down states
dominate, which suggests that indeed a spin polarization inversion takes place,
but at an insignificant magnitude. Interestingly, around 1.5 eV below the Fermi energy,
the spin up $d_{3z^{2}-r^2}$ peak of Co (below the thiophene) echoes
the spin up S peak at the same energy. In addition,
the spin down $d_{3z^{2}-r^2}$ peak of Co reappears in
the spin down density of states of the C1 group. Both peaks do not appear for the C2 group.

\begin{figure}[t]
\centering
\includegraphics[clip,width=0.5\textwidth]{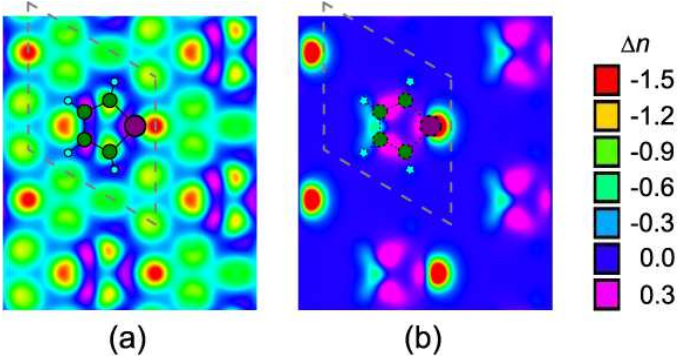}
\caption{(color online). Local spin polarization $\Delta\text{n}$
($10^{-3}$ $\text{e}/\text{bohr}^3$) of electrons in the energy
range $[E_{F}$$-0.4$ eV$,E_{F}]$, defined as the difference
between the spin up and spin down electron densities, at (a) the
interface between the substrate and the thiophene and (b) the
surface above the thiophene.} \label{fig:inversion}
\end{figure}

We plot in Fig.\ \ref{fig:inversion} the local spin polarization
of electrons in the energy range $[E_{F}$$-0.4$ eV$,E_{F}]$ at the
\ti{interface} and \ti{surface}, which constitutes the main
results of this Letter. The interface is defined as a plane
located between the thiophene molecule and the Co surface, about
0.9 \AA\ below the molecule. The surface is defined as a plane 0.9
\AA\ above the molecule. The spin polarizations at the interface
and surface have direct implications for the spin transport across
the interface and for scanning tunneling microscopy. Figure
\ref{fig:inversion} reveals an interesting polarization inversion:
at the interface (Fig.\ \ref{fig:inversion}(a)) the two C atoms at
the 2- and 5-positions (C2 group) are polarized in the same
direction as the substrate, while the two C atoms at the 3- and
4-positions (C1 group) are polarized in the opposite direction.
This reflects a polarization inversion, yet with a rather small
magnitude. At the surface (Fig.\ \ref{fig:inversion}(b)) we find
for the two C2 atoms barely any polarization, while for the two C1
atoms a small reversed polarization appears.

As a matter of fact, our results suggest that the $p_{z}$-$d$
exchange interaction proposed in Ref.\ [\onlinecite{atodiresei-prl-2010}]
is a viable mechanism to induce polarization at a spinterface,
but it is the detailed structure of the
molecule that determines the polarization inversion scheme.
To explain the peculiar polarization shown in Fig.\ \ref{fig:inversion},
we first need to address the effect of a reduced symmetry.
In the present case of thiophene on Co, the substitution of a C atom (from the \Cp\ molecule)
by an S atom breaks the symmetry that is maintained in a
$\pi$ electron system consisting of only C and H, thus naturally dividing the
C atoms into C1 and C2 groups.
The magnitude of the $p_{z}$-$d$ exchange coupling depends critically
on the overlap between the $p_{z}$ and $d$ orbitals.
In the case of \Cp\ on Fe, all five C atoms feel more or less
the same orbital overlap. We therefore expect a rather uniform spatial
distribution of the spin polarization. By its
spatially more extended orbitals, a S atom interacts with the C1 and C2 atoms differently.
The $p_{z}$-$d$ bonding is more distorted for the C2 atoms than for
the C1 atoms, rendering a much smaller $p_{z}$-$d$ overlap on C2 and,
thus, suppressing the inversion effect due to the $p_{z}$-$d$ exchange.
Being far from the S atom, the C1 atoms enjoy larger orbital overlaps,
leading to larger exchange coupling and polarization inversion.
It is difficult to estimate the magnitude of the exchange coupling in
the present setup. However, our results and qualitative picture based on symmetry
suggest that the exchange coupling in the thiophene on Co system is smaller than
what has been obtained for \Cp\ on Fe \cite{atodiresei-prl-2010}.
The results presented in this Letter reflect an intriguing
picture in terms of microscopic spin transmission through the interface:
depending on the route through which an electronic spin enters
the molecule, it either confronts a parallel (through C2 group) or
antiparallel alignment (through C1 group) of the local moment.

In conclusion, we have performed \ti{ab-initio} calculations
to investigate the spin polarization inversion at the interface between a thiophene
molecule and cobalt surface. In the thiophene molecule,
we have obtained a spin polarization inversion for the two C atoms
located far from the S atom, while neighboring C atoms show no inversion.
Our first principles results and symmetry arguments demonstrate that, although the
$p_{z}$-$d$ Zener exchange qualitatively explains the spin polarization
inversion, the viability of such a mechanism is affected by the detailed structure
of the organic molecule. For a systematic application of spinterfaces it
therefore is necessary to take into account the microscopic properties that govern
the polarization scheme, which can be highly inhomogeneous. The results obtained
in our work for a specific interface can be expected to be of general validity.

\end{document}